\documentclass[%
 preprint,
 showpacs,
 showkeys,
 preprintnumbers,
 amsmath,amssymb,
 aps,
  pra,
  longbibliography,
 ]{revtex4-1}

\usepackage{tikz}
\usepackage[breaklinks=true,colorlinks=true,anchorcolor=blue,citecolor=blue,filecolor=blue,menucolor=blue,pagecolor=blue,urlcolor=blue,linkcolor=blue]{hyperref}
\usepackage{graphicx}
\usepackage{url}

\usepackage{xcolor}

\begin{document}

\title{Dimensional lifting through generalized Gram--Schmidt process}


\author{Hans Havlicek}
\affiliation{Institute of Discrete Mathematics and Geometry, Vienna
    University of Technology, Wiedner Hauptstra\ss e 8-10/104, A-1040
    Vienna, Austria}
\email{havlicek@geometrie.tuwien.ac.at}
\homepage[]{http://www.geometrie.tuwien.ac.at/havlicek}

\author{Karl Svozil}
\affiliation{Institute for Theoretical Physics, Vienna
    University of Technology, Wiedner Hauptstra\ss e 8-10/136, A-1040
    Vienna, Austria}
\affiliation{Department of Computer Science, University of Auckland, Private Bag 92019,  Auckland 1142, New Zealand}
\email{svozil@tuwien.ac.at} \homepage[]{http://tph.tuwien.ac.at/~svozil}

\pacs{03.65.Aa,02.10.Ud,02.30.Sa,03.67.Ac}
\keywords{Orthogonality, quantum computation, Gram--Schmidt process}

\begin{abstract}
A new way of orthogonalizing ensembles of vectors by ``lifting'' them to higher dimensions is introduced. This method can potentially be utilized for solving quantum decision and computing problems.
\end{abstract}

\maketitle


The celebrated Gram-Schmidt algorithm allows the construction of a system of \emph{orthonormal}
vectors from an (ordered) system of \emph{linearly independent} vectors.
Let us mention that there exist a wide variety of proposals to ``generalize'' the Gram-Schmidt process~\cite{Leon-MR3054730} serving many different purposes.
In contrast to these generalizations, we construct a system of  \emph{orthogonal vectors}
from an (ordered) system of \emph{arbitrary} vectors, which may be linearly dependent.
(Even repeated vectors are allowed.) This task is accomplished by what will be called ``dimensional lifting.''

Some quantum computation tasks require the ortho\-gonalization of
previously non-orthogonal vectors. This might be best understood in terms of
mutually exclusive outcomes of generalized beam splitter experiments, where the
entire array of output ports corresponds to an ensemble of mutually orthogonal
subspaces, or, equivalently, mutually orthogonal perpendicular projection
operators~\cite{svozil-2016-vector}.

Of course, by definition (we may define a unitary transformation in a complex
Hilbert space by the requirement that it preserves the scalar
product~\cite[\S~73]{halmos-vs}), any transformation or mapping of
non-orthogonal vectors into mutually orthogonal ones will be non-unitary.
Yet we may resort to requiring
that some sort of angles or distances (e.g., in the original Hilbert space)
remain unchanged.


Suppose, for the sake of demonstration, two non-orthogonal vectors, and suppose
further that somehow one could ``orthogonalize'' them while at the same time
retaining structural elements, such as the angles
between projections of the new, mutually orthogonal vectors
onto the subspace spanned by the original vectors.
For instance, the two non-orthogonal vectors could be transformed into
vectors of some higher-dimensional Hilbert space satisfying the following
properties with respect to the original vectors: (i) the new vectors are
orthogonal, and (ii) the orthogonal projection along the new, extra
dimension(s) of the two vectors render the original vectors. A straightforward
three-dimensional construction with the desired outcome can be given as
follows: suppose the original vectors are unit vectors denoted by $\vert {\bf
e}_1 \rangle$ and $\vert {\bf e}_2 \rangle$; and $0 < \vert \langle {\bf e}_1
\vert {\bf e}_2 \rangle \vert < 1$. Suppose further a two-dimensional
coordinate frame in which $\vert {\bf e}_1 \rangle$ and $\vert {\bf e}_2
\rangle$ are planar; thus we can write, in terms of some orthonormal basis
$\vert {\bf e}_1 \rangle =\begin{pmatrix}x_{1,1},x_{1,2}\end{pmatrix}$ as well
as $\vert {\bf e}_2 \rangle =\begin{pmatrix}x_{2,1},x_{2,2}\end{pmatrix}$.
Suppose we ``enlarge'' the vector space to include an additional dimension, and
suppose a Cartesian basis system in that greater space which includes the two
vectors of the old basis (and an additional unit vector which is orthogonal
with respect to the original plane spanned by the original basis vectors).

{\it Ad hoc} it is rather intuitive how two (not necessarily unit) vectors can
be found which project onto the original vectors, and which are orthogonal:
``create'' a three-dimensional vector space with one extra dimension, assign
a non-zero extra coordinate (such as $1$) associated with this dimension for
the first vector, and use the extra coordinate of the second vector for
compensate any nonzero value of the scalar product of the two original vectors;
in particular, whose coordinates with respect to the new basis are
\begin{equation}
\begin{aligned}
\vert {\bf f}_1 \rangle =& \begin{pmatrix}x_{1,1},x_{1,2},1\end{pmatrix},\\
\vert {\bf f}_2 \rangle =&
\begin{pmatrix}
x_{2,1},x_{2,2}, - \left( x_{1,1}x_{2,1} + x_{1,2}x_{2,2} \right)
\end{pmatrix},
\end{aligned}
\end{equation}
which are orthogonal by construction.


It is not too difficult to find explicit constructions for the more general
case of $k$ vectors $\vert {\bf e}_1 \rangle , \ldots, \vert {\bf e}_k \rangle
$ in ${\Bbb R}^n$ (cf. Ref.~\cite{svozil-2016-vector} for a rather inefficient
method).

In the following, for the sake of construction, we shall embed ${\Bbb R}^n$
into ${\Bbb R}^{n+k}$, such that we fill all additional vector coordinates of
$\vert {\bf e}_1 \rangle , \ldots \vert {\bf e}_k \rangle $ with zeroes. For
the new, mutually orthogonal, vectors we make the following {\it Ansatz} by
defining
\begin{equation}\label{ansatz}
\begin{aligned}
\vert {\bf f}_1 \rangle =& \begin{pmatrix}{\bf e}_1, 1, 0, \ldots, 0 \end{pmatrix}, \\
\vert {\bf f}_2 \rangle =& \begin{pmatrix}{\bf e}_2, x_{2,1}, 1, 0,\ldots, 0\end{pmatrix},\\
\ldots& \\
\vert {\bf f}_k \rangle =& \begin{pmatrix}{\bf e}_k, x_{k,1}, x_{k,2},\ldots, x_{k,k-1}, 1\end{pmatrix},
\end{aligned}
\end{equation}
with yet to be determined coordinates $x_{i,j}$.
(The symbols ${\bf e}_i$ stand for all the $n$ coordinates of $\vert {\bf e}_1 \rangle$.)

The unit coordinates $1$ ensure that the new vectors are linearly
independent. By construction the orthogonal projection of $\vert {\bf f}_i
\rangle$ onto ${\Bbb R}^n$ renders $\vert {\bf e}_i \rangle$ for all $1\le i
\le k$.

What remains is the recursive determination of the unknown coordinates $x_{i,j}$.
Note that all  $\vert {\bf f}_j \rangle$ must satisfy the following relations:
for $j > 1$,
orthogonality demands that
$\langle {\bf f}_1 \vert {\bf f}_j \rangle = 0$, and therefore
$\langle {\bf e}_1 \vert {\bf e}_j \rangle + 1 \cdot  x_{j,1} = 0$,
and therefore
\begin{equation}
x_{j,1} = - \langle {\bf e}_1 \vert {\bf e}_j \rangle
.
\end{equation}
In this way all unknown coordinates $x_{2,1}, \ldots , x_{k,1}$ can be determined.

Similar constructions yield the remaining unknown coordinates in
$\vert {\bf f}_2 \rangle ,
\ldots       ,
\vert {\bf f}_k \rangle$.
For $j > 2$, $\langle {\bf f}_2 \vert {\bf f}_j \rangle = 0$,
and therefore
$
\langle {\bf e}_2 \vert {\bf e}_j \rangle + x_{2,1}  x_{j,1} +  x_{j,2} = 0
$,
yielding
\begin{equation}
x_{j,2} =  - \langle {\bf e}_2 \vert {\bf e}_j \rangle - x_{2,1}  x_{j,1}.
\end{equation}
In this way all unknown coordinates $x_{3,2}, \ldots , x_{k,2}$ can be determined.

This procedure is repeated until one arrives at $j = k-1$, and therefore at the orthogonality of
$\vert {\bf f}_{k-1} \rangle
$ and $
\vert {\bf f}_k \rangle$, encoded by the condition
$\langle {\bf f}_{k-1} \vert {\bf f}_k \rangle = 0$, and hence
\begin{equation}
\begin{aligned}
x_{k,k-1} = - \left( \langle {\bf e}_{k-1} \vert {\bf e}_k \rangle + \right.
\\
\left. + x_{k-1,1}  x_{k,1} + \cdots + x_{k-1,k-2}  x_{k,k-2} \right).
\end{aligned}
\end{equation}

The approach has the advantage that, at each stage of the recursive
construction, there is only a single unknown coordinate per equation.
This situation is well known from Gaussian elimination.
The {\it Ansatz}
also works if one of the original vectors is the zero vector, and if some of
the original vectors are equal.

The resulting system of orthogonal vectors is not the only solution of the initial problem
-- to find orthogonal vector which project onto the original ones -- which can be
explicitly demonstrated by multiplying all vectors $\vert {\bf f}_1 \rangle ,
\ldots , \vert {\bf f}_k \rangle$ with the matrix
\begin{equation}
{\rm diag} \begin{pmatrix} {\Bbb I}_n,c \textsf{\textbf{T}}\end{pmatrix},
\end{equation}
whereby ${\Bbb I}_n$ stands for the $n$-dimensional unit matrix, $c$ can be a
real nonzero constant, and $\textsf{\textbf{T}}$ is a $k$-dimensional
orthogonal matrix. (For complex Hilbert space, the orthogonal matrix needs to
be substituted by a unitary matrix, and by a complex constant $c\neq 0$.)

On the other hand, we may reinterpret our procedure as follows:
Let
$\vert{\bf e}_1\rangle, \ldots,\vert{\bf e}_k\rangle$ be a system of vectors
in ${\Bbb R}^n$, not necessarily spanning ${\Bbb R}^n$, and not necessarily being linearly independent.
(The ordering of the vectors in this system will be essential throughout.)
We embed ${\Bbb R}^n$ in ${\Bbb R}^{n+k}$ as we did above and
denote the orthogonal complement of ${\Bbb R}^n$ by $C \cong {\Bbb R}^k$.
Therefore, the first $n$ coordinates of all  vectors in $C$  vanish,
and ${\Bbb R}^{n+k}$ can be represented by a direct sum
${\Bbb R}^{n+k}= {\Bbb R}^n \oplus {\Bbb R}^k$.
Additionally, we choose some (ordered) orthonormal basis of $C$, say,
$\vert{\bf g}_{1}\rangle, \ldots,\vert{\bf g}_{k}\rangle$.

Then there is a {\em unique} system
of orthogonal vectors $\vert{\bf f}_1\rangle, \ldots,\vert{\bf f}_{k}\rangle$
in ${\Bbb R}^{n+k}$ such that the following conditions are satisfied:
\begin{enumerate}
  \item\label{cond1} For all $1\leq i\leq k$ the orthogonal projection of
      ${\Bbb R}^{n+k}$ onto ${\Bbb R}^{n}$ sends $\vert{\bf f}_i\rangle$ to
      $\vert{\bf e}_{i}\rangle$.

  \item\label{cond2} The orthogonal projection of ${\Bbb R}^{n+k}$ onto $C$
      sends $\vert{\bf f}_1\rangle,\ldots,\vert{\bf f}_k\rangle$ to some
      (ordered) basis of the subspace $C$. Applying the Gram-Schmidt
      process to this (ordered) basis gives the orthonormal basis
      $\vert{\bf g}_{1}\rangle,\ldots,\vert{\bf g}_{k}\rangle$.
\end{enumerate}

Indeed, in our previous \emph{Ansatz} we tacitly assumed the orthonormal basis
$\vert{\bf g}_{1}\rangle, \ldots,\vert{\bf g}_{k}\rangle$
of $C$ to comprise the orthogonal projections of
the last $k$ vectors of the standard basis
$\vert{\bf b}_1\rangle,\ldots,\vert{\bf b}_{n+k}\rangle$ of ${\Bbb R}^{n+k}$ onto $C$.
Condition
\ref{cond2} enforces the presence of
all the $1$'s and $0$'s in formula \eqref{ansatz}, since
the Gram-Schmidt process, applied to the vectors
\begin{equation*}
    \vert{\bf f}_1\rangle-\vert{\bf e}_1\rangle,\ldots,\vert{\bf f}_k\rangle-\vert{\bf e}_k\rangle
,
\end{equation*}
has to result in
$\vert{\bf b}_{n+1}\rangle,\ldots,\vert{\bf b}_{n+k}\rangle$.
Notice that the usual Gram-Schmidt process gives merely an \emph{orthogonal\/}
basis, whose vectors can be normalized in a second step in order to obtain an
\emph{orthonormal\/} basis. In our setting, however, such a second step is not
allowed. As we saw above, now Condition \ref{cond1} guarantees that
$\vert{\bf f}_1\rangle, \ldots,\vert{\bf f}_{k}\rangle$
are uniquely determined.

Besides uniqueness, this construction has the additional advantage
that the dot product in ${\Bbb R}^{n+k}$ ``decays''
into the sum of dot products in ${\Bbb R}^{n}$
and in ${\Bbb R}^{k}$:
any basis vector ${\bf f}_i \in {\Bbb R}^{n+k}$
can be uniquely written as  ${\bf f}_i = {\bf e}_i + {\bf h}_i$,
where ${\bf e}_i$ and ${\bf h}_i$
represent
the projection of ${\bf f}_i$ along ${\bf h}_i$ onto the original subspace ${\Bbb R}^{n}$,
and
the projection of ${\bf f}_i$ along ${\bf e}_i$ onto $C$,
respectively.
Since ${\bf e}_i$ is orthogonal to ${\bf h}_i$,
for $i\neq j$,
$
{\bf f}_i \cdot {\bf f}_j
=   {\bf e}_i \cdot {\bf e}_j + {\bf h}_i \cdot  {\bf h}_j = 0,
$
and thus
\begin{equation}
\begin{aligned}
{\bf e}_i \cdot {\bf e}_j = -{\bf h}_i \cdot  {\bf h}_j
.
\end{aligned}
\end{equation}


Let us, for the sake of a physical example,
study configurations associated with decision problems which can be efficiently
(that is, with some speedup with respect to purly classical means~\cite{svozil-2016-vector}) encoded quantum mechanically.
The {\em inverse problem} is the projection of orthogonal systems of vectors onto lower dimensions.
This method renders a system of non-orthogonal rays, also called {\it eutactic stars}~\cite{schlaefli-1901,hadwiger-40,coxeter-polytopes,seidel-76,hasse-stachel96}
which can be effectively levied to mutually exclusive outcomes in
generalized beam splitter
configurations~\cite{rzbb,zukowski-97}
reflecting the higher dimensional Hilbert space.

One instance of such a quantum computation involving the reduction to ensembles of orthogonal vectors (and their associated
span or projection operators) is the Deutsch-Jozsa algorithm, as reviewed in Ref.~\cite{svozil-2016-vector}.
Another, somewhat contrived, problem can be constructed in three dimensions from an eutactic star
\begin{equation}
\begin{aligned}
 \frac{1}{\sqrt{3}} \left\{
 \begin{pmatrix}1,1\end{pmatrix},
 \begin{pmatrix}\frac{1}{2} \left[{\sqrt{3}} i-1\right] , \frac{1}{2} \left[-{\sqrt{3}} i-1\right] \end{pmatrix}, \right. \\
\left.
\begin{pmatrix}  \frac{1}{ 2} \left[-{\sqrt{3}} i-1\right] , \frac{1}{ 2} \left[{\sqrt{3}} i-1\right]  \end{pmatrix}
   \right\} ,
\end{aligned}
\end{equation}
which is the projection onto the plane formed by the first two coordinates
of a three-dimensional orthormal basis
\begin{equation}
\begin{aligned}
   {\mathfrak B}_3 = \frac{1}{\sqrt{3}} \left\{
 \begin{pmatrix}1,1,1\end{pmatrix},\right. \\
 \begin{pmatrix}\frac{1}{2} \left[{\sqrt{3}} i-1\right] , \frac{1}{2} \left[-{\sqrt{3}} i-1\right] ,
  1\end{pmatrix}, \\
\left.
\begin{pmatrix}
 \frac{1}{ 2} \left[-{\sqrt{3}} i-1\right] , \frac{1}{ 2} \left[{\sqrt{3}} i-1\right] , 1  \end{pmatrix}
   \right\} ,
\end{aligned}
\end{equation}
which, together with the Cartesian standard basis, forms a pair of unbiased bases~\cite{Schwinger.60}.

Still another decision configuration is the eutactic star
\begin{equation}
\begin{aligned}
\frac{1}{2}\left\{
 \begin{pmatrix}1,1,1\end{pmatrix},
 \begin{pmatrix}1,1,-1\end{pmatrix}, \right.\\
 \begin{pmatrix}1,-1,1\end{pmatrix},
\left.
 \begin{pmatrix}1,-1,-1\end{pmatrix}
   \right\} ,
\end{aligned}
\end{equation}
which is the projection onto the subspace formed by the first three coordinates
of a four-dimensional orthormal basis
\begin{equation}
\begin{aligned}
{\mathfrak B}_4 =\frac{1}{2} \left\{
 \begin{pmatrix}1,1,1,1\end{pmatrix},
 \begin{pmatrix}1,1,-1,-1\end{pmatrix}, \right.\\
\left.
 \begin{pmatrix}1,-1,1,-1\end{pmatrix},
 \begin{pmatrix}1,-1,-1,1\end{pmatrix}
   \right\} .
\end{aligned}
\end{equation}

More concretely, suppose some, admittedly construed, function $f$, and some quantum encoding $\vert x f(y) \rangle$,
where $x$ and $y$ stand for (sequences of) auxiliary and input bits, respectively,
would yield one of the basis systems
${\mathfrak B}_3$ or
${\mathfrak B}_4$.
By reducing the auxiliary bits $x$, one might end up with the
eutactic stars introduced above.
Alas, so far, no candidate of this kind has been proposed.


In summary, a new method of orthogonalizing ensembles of vectors
has been introduced.
Thereby, the original vectors are  ``lifted'' to
or ``completed'' in higher dimensions.
This method could be utilized for solving quantum decision and computing problems
if the original problem does not allow an orthogonal encoding,
and if extra bits can be introduced which render the equivalent of the extra dimensions
in which the original state vectors can be lifted and orthogonalized.

Compared with  methods which were introduced~\cite{IVANOVIC1987257,PERES198819,JAEGER199583} previously to optimally differentiate between two non-orthogonal states,
the scheme suggested here is similar in the sense that, in order to obtain a better resolution, the effective dimensionality of the problem is increased.
However, our scheme is not limited to the differentiation between two states, as it uses arbitrary dimensionality.
More importantly, whereas our scheme is capable of separating different states precisely, but in general is non-unitarity
(indeed, the original vectors are not mutually orthogonal but the lifted vector are, thereby changing the angles among vectors;
resulting in transformations that cannot be unitary),
the former method is only probabilistic but unitary.

\begin{acknowledgments}
This work was supported in part by the European Union, Research Executive Agency (REA),
Marie Curie FP7-PEOPLE-2010-IRSES-269151-RANPHYS grant.
\end{acknowledgments}


%

\end{document}